\documentclass{PoS}

\usepackage{graphicx}
\usepackage{psfrag}
\usepackage{graphics}
\usepackage{float}
\usepackage{amssymb,epsfig,amsmath,euscript,array}
\usepackage{cite}
\usepackage{multicol}
\usepackage{subfig}
\usepackage{fancyhdr}
\usepackage{color}
\usepackage{scalerel,stackengine}

\newcommand{\be}{\begin{equation}}
\newcommand{\ee}{\end{equation}}

\newcommand{\bm}[1]{\mbox{\boldmath $#1$}}

\def\bd{\begin{document}}
\def\ed{\end{document}}
\def\nn{\nonumber}
\def\bea{\begin{eqnarray}}
\def\eea{\end{eqnarray}}
\let\bm=\bibitem
\let\la=\label

\def\npb#1#2#3{Nucl. Phys. {\bf{B#1}} #3 (#2)}
\def\plb#1#2#3{Phys. Lett. {\bf{#1B}} #3 (#2)}
\def\prl#1#2#3{Phys. Rev. Lett. {\bf{#1}} #3 (#2)}
\def\prd#1#2#3{Phys. Rev. {D \bf{#1}} #3 (#2)}
\def\cmp#1#2#3{Comm. Math. Phys. {\bf{#1}} #3 (#2)}
\def\cqg#1#2#3{Class. Quantum Grav. {\bf{#1}} #3 (#2)}
\def\nppsa#1#2#3{Nucl. Phys. B (Proc. Suppl.) {\bf{#1A}}#3 (#2)}
\def\ap#1#2#3{Ann. of Phys. {\bf{#1}} #3 (#2)}
\def\ijmp#1#2#3{Int. J. Mod. Phys. {\bf{A#1}} #3 (#2)}
\def\rmp#1#2#3{Rev. Mod. Phys. {\bf{#1}} #3 (#2)}
\def\mpla#1#2#3{Mod. Phys. Lett. {\bf A#1} #3 (#2)}
\def\jhep#1#2#3{J. High Energy Phys. {\bf #1} #3 (#2)}
\def\atmp#1#2#3{Adv. Theor. Math. Phys. {\bf #1} #3 (#2)}

\newcommand{\EQ}[1]{\begin{equation} #1 \end{equation}}
\newcommand{\AL}[1]{\begin{subequations}\begin{align} #1 \end{align}\end{subequations}}
\newcommand{\SP}[1]{\begin{equation}\begin{split} #1 \end{split}\end{equation}}
\newcommand{\ALAT}[2]{\begin{subequations}\begin{alignat}{#1} #2 \end{alignat}\end{subequations}}
\def\beq{\begin{equation}}
\def\eeq{\end{equation}}

\def\N{{\cal N}}
\def\sst{\scriptscriptstyle}
\def\thetabar{\bar\theta}
\def\Tr{{\rm Tr}}
\def\one{\mbox{1 \kern-.59em {\rm l}}}
 \def\Nh{\hat{N}}

\def\a{\alpha}      \def\da{{\dot\alpha}}
\def\b{\beta}       \def\db{{\dot\beta}}
\def\c{\gamma}  \def\G{\Gamma}  \def\cdt{\dot\gamma}
\def\d{\delta}  \def\D{\Delta}  \def\ddt{\dot\delta}
\def\e{\epsilon}        \def\vare{\varepsilon}
\def\f{\phi}    \def\F{\Phi}    \def\vvf{\f}
\def\h{\eta}
\def\k{\kappa}
\def\l{\lambda} \def\L{\Lambda}
\def\m{\mu} \def\n{\nu}
\def\o{\omega}
\def\p{\pi} \def\P{\Pi}
\def\r{\rho}
\def\s{\sigma}  \def\S{\Sigma}
\def\t{\tau}
\def\th{\theta} \def\Th{\Theta} \def\vth{\vartheta}
\def\X{\Xeta}
\def\z{\zeta}

\def\cA{{\cal A}} \def\cB{{\cal B}} \def\cC{{\cal C}}
\def\cD{{\cal D}} \def\cE{{\cal E}} \def\cF{{\cal F}}
\def\cG{{\cal G}} \def\cH{{\cal H}} \def\cI{{\cal I}}
\def\cJ{{\cal J}} \def\cK{{\cal K}} \def\cL{{\cal L}}
\def\cM{{\cal M}} \def\cN{{\cal N}} \def\cO{{\cal O}}
\def\cP{{\cal P}} \def\cQ{{\cal Q}} \def\cR{{\cal R}}
\def\cS{{\cal S}} \def\cT{{\cal T}} \def\cU{{\cal U}}
\def\cV{{\cal V}} \def\cW{{\cal W}} \def\cX{{\cal X}}
\def\cY{{\cal Y}} \def\cZ{{\cal Z}}

\def\ua{\underline{\alpha}}
\def\ub{\underline{\phantom{\alpha}}\!\!\!\beta}
\def\uc{\underline{\phantom{\alpha}}\!\!\!\gamma}
\def\um{\underline{\mu}}
\def\ud{\underline\delta}
\def\ue{\underline\epsilon}
\def\una{\underline a}\def\unA{\underline A}
\def\unb{\underline b}\def\unB{\underline B}
\def\unc{\underline c}\def\unC{\underline C}
\def\und{\underline d}\def\unD{\underline D}
\def\une{\underline e}\def\unE{\underline E}
\def\unf{\underline{\phantom{e}}\!\!\!\! f}\def\unF{\underline F}
\def\unm{\underline m}\def\unM{\underline M}
\def\unn{\underline n}\def\unN{\underline N}
\def\unp{\underline{\phantom{a}}\!\!\! p}\def\unP{\underline P}
\def\unq{\underline{\phantom{a}}\!\!\! q}
\def\unQ{\underline{\phantom{A}}\!\!\!\! Q}
\def\unH{\underline{H}}

\def\As {{A \hspace{-6.4pt} \slash}\;}
\def\bs {{b \hspace{-6.4pt} \slash}\;}
\def\Ds {{D \hspace{-6.4pt} \slash}\;}
\def\ds {{\del \hspace{-6.4pt} \slash}\;}
\def\ss {{\s \hspace{-6.4pt} \slash}\;}
\def\ks {{ k \hspace{-6.4pt} \slash}\;}
\def\ps {{p \hspace{-6.4pt} \slash}\;}
\def\pas {{{p_1} \hspace{-6.4pt} \slash}\;}
\def\pbs {{{p_2} \hspace{-6.4pt} \slash}\;}

\def\Fh{\hat{F}}
\def\Vh{\hat{V}}
\def\Xh{\hat{X}}
\def\ah{\hat{a}}
\def\xh{\hat{x}}
\def\yh{\hat{y}}
\def\ph{\hat{p}}
\def\xih{\hat{\xi}}

\def\psit{\tilde{\psi}}
\def\Psit{\tilde{\Psi}}
\def\tht{\tilde{\th}}
\def\lt{\tilde{\lambda}}
\def\At{\tilde{A}}
\def\Qt{\tilde{Q}}
\def\Rt{\tilde{R}}
\def\Nt{\tilde{N}}

\def\at{\tilde{a}}
\def\st{\tilde{s}}
\def\ft{\tilde{f}}
\def\pt{\tilde{p}}
\def\qt{\tilde{q}}
\def\vt{\tilde{v}}
\def\nt{\tilde{n}}

\def\delb{\bar{\partial}}
\def\bz{\bar{z}}
\def\bD{\bar{D}}
\def\bB{\bar{B}}

\def\bk{{\bf k}}
\def\bl{{\bf l}}
\def\bp{{\bf p}}
\def\bq{{\bf q}}
\def\br{{\bf r}}
\def\bx{{\bf x}}
\def\by{{\bf y}}
\def\bR{{\bf R}}
\def\bV{{\bf V}}

\def\d{\delta}\def\D{\Delta}\def\ddt{\dot\delta}

\def\pa{\partial} \def\del{\partial}
\def\xx{\times}
\def\uno{\mbox{1 \kern-.59em {\rm l}}}

\def\trp{^{\top}}
\def\inv{^{-1}}
\def\dag{{^{\dagger}}}
\def\pr{^{\prime}}
\def\lan{\langle}
\def\ran{\rangle}

\def\rar{\rightarrow}
\def\lar{\leftarrow}
\def\lrar{\leftrightarrow}

\newcommand{\0}{\,\!}      %this is just NOTHING!
\def\one{1\!\!1\,\,}
\def\im{\imath}
\def\jm{\jmath}

\newcommand{\tr}{\mbox{tr}}
\newcommand{\slsh}[1]{/ \!\!\!\! #1}

\def\vac{|0\rangle}
\def\lvac{\langle 0|}

\def\hlf{\frac{1}{2}}
\def\ove#1{\frac{1}{#1}}

\def\Box{\square}
\def\ZZ{\mathbb{Z}}
\def\CC#1{({\bf #1})}
\def\bcomment#1{}
\def\bfhat#1{{\bf \hat{#1}}}
\def\VEV#1{\left\langle #1\right\rangle}

\newcommand{\ex}[1]{{\rm e}^{#1}} \def\ii{{\rm i}}

\def\rr{{\rm r}} \def\rs{{\rm s}}\def\rv{{\rm v}}
\def\ri{{\rm i}}\def\rj{{\rm j}}

\newcommand{\lrbrk}[1]{\left(#1\right)}
\newcommand{\sfrac}[2]{{\textstyle\frac{#1}{#2}}}

\newcommand\equalhat{\mathrel{\stackon[1.5pt]{=}{\stretchto{%
    \scalerel*[\widthof{=}]{\wedge}{\rule{1ex}{3ex}}}{0.5ex}}}}

\font\mybb=msbm10 at 12pt
\def\bb#1{\hbox{\mybb#1}}

\font\myBB=msbm10 at 18pt
\def\BB#1{\hbox{\myBB#1}}

\newcommand{\tclr}{\textcolor}
\newcommand{\bpmat}{\begin{pmatrix}}
\newcommand{\epmat}{\end{pmatrix}}
\newcommand{\mrm}[1]{\mathrm{#1}}
\newcommand{\mrs}[1]{\scriptscriptstyle{\mathrm{#1}}}
\newcommand{\vct}[1]{\boldsymbol{#1}}
\newcommand{\hf}{\frac{1}{2}}
\newcommand{\x}{\times}
\newcommand{\pd}{\partial}
\newcommand{\dslash}{\displaystyle{\not}}
\newcommand{\ol}[1]{\overline{#1}}
\newcommand{\abs}[1]{\vert{#1}\vert}
\newcommand{\chiSqM}{\chi^2_{\mrm{min}}}
\newcommand{\chiSqMDof}{\chi^2_{\mrm{min}}/\mrm{d.o.f.}}
\newcommand{\om}{\omega}
\newcommand{\Lag}{\mathcal{L}}
\newcommand{\ord}{\mathcal{O}}
\newcommand{\eps}{\epsilon}
\newcommand{\beFrac}{\frac{1-\be}{1+\be}}
\newcommand{\beFracI}{\frac{1+\be}{1-\be}}
\newcommand{\amu}{a_{\mu}}
\newcommand{\damu}{\delta\amu}
\newcommand{\Damu}{\Delta\amu}
\newcommand{\amuUnit}{10^{-10}}
\newcommand{\mmu}{m_{\mu}}
\newcommand{\amuQED}{\amu^{\mrm{QED}}}
\newcommand{\amuEW}{\amu^{\mrm{EW}}}
\newcommand{\amuEWl}{\amu^{\mrm{EW,}\,1l}}
\newcommand{\amuEWll}{\amu^{\mrm{EW,}\,2l}}
\newcommand{\amuh}{\amu^{\mrm{had}}}
\newcommand{\amuhLO}{\amu^{\text{had, LOVP}}}
\newcommand{\amuhHO}{\amu^{\text{had, HOVP}}}
\newcommand{\amuhHOa}{\amu^{\text{had, HOVP(a)}}}
\newcommand{\amuhHOb}{\amu^{\text{had, HOVP(b)}}}
\newcommand{\amuhHOc}{\amu^{\text{had, HOVP(c)}}}
\newcommand{\amuhLbL}{\amu^{\text{had, LbL}}}
\newcommand{\ff}[3]{\mathcal{F}_{\pi^{0{#1}}\gamma^{#2}\gamma^{#3}}}
\newcommand{\alps}{\alpha_s}
\newcommand{\asmz}{\alpha_s(M_Z^2)}
\newcommand{\amz}{\alpha(M_Z^2)}
\newcommand{\aqmz}{\alpha_{\mrm{QED}}(M_Z^2)}
\newcommand{\delAlp}{\Delta\alpha}
\newcommand{\dAlpL}{\delAlp_{\mrm{lep}}}
\newcommand{\dAlpT}{\delAlp_{\mrm{top}}}
\newcommand{\dAlpH}{\delAlp_{\mrm{had}}}
\newcommand{\dAlpHF}{\dAlpH^{(5)}}
\newcommand{\dAlpHFmz}{\dAlpHF(M_Z^2)}
\newcommand{\tmin}{t_{\mrm{min}}}
\newcommand{\sTh}{s_{\mrm{th}}}
\newcommand{\eTh}{\sqrt{\sTh}}
\newcommand{\Ekmi}{E^{\,(k,m)}_i}
\newcommand{\Nkm}{N^{(k,m)}}
\newcommand{\Nkn}{N^{(k,n)}}
\newcommand{\Nexp}{N_{\mrm{exp}}}
\newcommand{\Nclu}{N_{\mrm{clu}}}
\newcommand{\Ntot}{N_{\mrm{tot}}}
\newcommand{\Rkmi}{R^{\,(k,m)}_i}
\newcommand{\Rknj}{R^{\,(k,n)}_j}
\newcommand{\dRkmi}{\mrm{d}\Rkmi}
\newcommand{\dRtkmi}{\mrm{d}\tilde{R}^{\,(k,m)}_i}
\newcommand{\BR}[2]{\mathcal{B}(#1\to #2)}
\newcommand{\decay}[2]{#1\to #2}
\newcommand{\UpsIVs}{\Upsilon(4S)}
\newcommand{\Gee}{\Gamma_{ee}}
\newcommand{\Gtot}{\Gamma_{\mrm{tot}}}
\newcommand{\ppC}{\pi^+\pi^-}
\newcommand{\ppN}{\pi^0\pi^0}
\newcommand{\pppC}{\pi^+\pi^-\pi^0}
\newcommand{\kkC}{K^+K^-}
\newcommand{\kskl}{K^0_S K^0_L}
\newcommand{\ksks}{K^0_S K^0_S}
\newcommand{\klkl}{K^0_L K^0_L}
\newcommand{\kskp}{K^0_S K^{\pm}\pi^{\mp}}
\newcommand{\eeMuMu}{e^+e^-\to\mu^+\mu^-}
\newcommand{\eeHadr}{e^+e^-\to\mrm{hadrons}}
\newcommand{\eeGhadr}{e^+e^-\to\gamma^*\to\mrm{hadrons}}
\newcommand{\tauNuHadr}{\tau\to\nu_{\tau}+\mrm{hadrons}}
\newcommand{\eeGPiPi}{e^+e^-\to\gamma^*\to\pi^+\pi^-}
\newcommand{\tauNuWNuPiPi}{\tau\to\nu_{\tau}W\to\nu_{\tau}\pi\pi^0}
\newcommand{\eeGIncl}{e^+e^-\to\gamma^*\to\mrm{all\,hadrons}}
\newcommand{\eeIncl}{e^+e^-\to\mrm{all\,hadrons}}
\newcommand{\eePiG}{e^+e^-\to\pi^0\gamma}
\newcommand{\eePiPi}{e^+e^-\to\pi^+\pi^-}
\newcommand{\eePiPiPi}{e^+e^-\to\pi^+\pi^-\pi^0}
\newcommand{\eeKK}{e^+e^-\to K^+K^-}
\newcommand{\ch}{\mrm{ch}}
\newcommand{\iso}{\mrm{iso}}
\newcommand{\noeta}{\text{no }\eta}
\newcommand{\kkr}{K\bar{K}\rho}
\newcommand{\kkp}{K\bar{K}\pi}
\newcommand{\kkpp}{K\bar{K}2\pi}
\newcommand{\kkppp}{K\bar{K}3\pi}
\newcommand{\isoAA}{(2\pi^+2\pi^-\pi^0)_{\mrm{no}\,\eta}}
\newcommand{\isoAB}{(\pi^+\pi^-3\pi^0)_{\mrm{no}\,\eta}}
\newcommand{\isoAC}{\omega(\to\mrm{npp})2\pi}
\newcommand{\isoACf}{\omega(\to\text{non-pure pionic states})2\pi}
\newcommand{\isoAD}{\eta\pi^+\pi^-}
\newcommand{\isoBA}{(2\pi^+2\pi^-2\pi^0)_{\mrm{no}\,\eta}}
\newcommand{\isoBB}{(\pi^+\pi^-4\pi^0)_{\mrm{no}\,\eta}}
\newcommand{\isoBC}{3\pi^+3\pi^-}
\newcommand{\isoBD}{\omega(\to\mrm{npp})3\pi}
\newcommand{\isoBDf}{\omega(\to\text{non-pure pionic state})3\pi}
\newcommand{\isoBE}{\eta\omega}
\newcommand{\isoEA}{\kkppp}
\newcommand{\isoEAa}{(K^+K^-\pi^+\pi^-\pi^0)_{\mrm{no}\,\eta}}
\newcommand{\isoEAb}{(K^0\bar{K}^0\pi^+\pi^-\pi^0)_{\mrm{no}\,\eta}}
\newcommand{\isoEB}{\omega(\to\mrm{npp})K\bar{K}}
\newcommand{\isoEBf}{\omega(\to\text{non-pure pionic states})K\bar{K}}
\newcommand{\isoEC}{\eta\phi}
\newcommand{\isoFA}{\eta2\pi^+2\pi^-}
\newcommand{\isoFB}{\eta\pi^+\pi^-2\pi^0}
\newcommand{\sigEEhadr}{\sigma(\eeHadr)}
\newcommand{\sigHad}{\sigma_{\mrm{had}}}
\newcommand{\sigHadB}{\sigHad^0}
\newcommand{\sigPt}{\sigma_{\mrm{pt}}}
\newcommand{\Rhad}{R_{\mrm{had}}}

\title{The muon $g-2$: a brief overview of hadronic cross section data}

\ShortTitle{The muon $g-2$: a brief overview of hadronic cross section data}

\author{\speaker{Alexander Keshavarzi}\thanks{The author would like to thank the organisers of {\em The 9th International Workshop on Chiral Dynamics} for a very productive and enjoyable workshop. Special thanks and acknowledgement are extended to Daisuke Nomura and Thomas Teubner for their collaboration on the KNT18 analysis. This manuscript has been authored by an employee of The University of Mississippi supported in-part by the U.S. Department of Energy Office of Science, Office of High Energy Physics, award DE-SC0012391 and using the resources of the Fermi National Accelerator Laboratory (Fermilab), a U. S. Department of Energy, Office of Science, HEP User Facility. The work of the author at the University of Liverpool was supported by STFC under the consolidated grant ST/N504130/1.}\\
        Department of Physics and Astronomy, The University of Mississippi, Mississippi 38677, U.S.
\\Department of Mathematical Sciences, University of Liverpool,
Liverpool L69 3BX, U.K. \\
        E-mail: \email{aikeshav@olemiss.edu}}

\abstract{
The hadronic vacuum polarisation contributions to the anomalous magnetic moment of the muon, $a_{\mu}^{\rm had, VP}$ are evaluated dispersively via a combination of experimentally measured $e^+e^-\rightarrow {\rm hadrons}$ cross section data. Many experiments have dedicated programmes to precisely measure these final states, meaning that a vast amount of data is now available and that, in some cases, overall precision has reached the sub-percent level. However, data tensions are evident between measurements of the same hadronic channels from different experiments, which reduces the overall quality of the data combinations used to determine $a_{\mu}^{\rm had, \, VP}$. The inclusion of these data in the KNT18 analysis results in $a_{\mu}^{\rm had, \, LO \, VP} = (693.26 \pm 2.46)\times 10^{-10}$ and $a_{\mu}^{\rm had, \, NLO \, VP} = (-9.82 \pm 0.04)\times 10^{-10}$. The corresponding new estimate for the Standard Model prediction is found to be $a_{\mu}^{\rm SM}  =  (11\ 659 \ 182.04  \pm 3.56) \times 10^{-10}$, which is $3.7\sigma$ below the current experimental measurement.
}

\FullConference{The 9th International Workshop on Chiral Dynamics\\
		17-21 September 2018\\
		Durham, NC, USA}

\begin{document}

\section{Introduction}

The anomalous magnetic moment of the muon, $a_{\mu} = (g-2)_{\mu}/2$,
stands as an enduring test of the Standard Model (SM), where the
$\sim3.5\sigma$  (or higher) discrepancy between the experimental
measurement $a_{\mu}^{\rm exp} = 11\ 659 \ 209.1 \ (5.4) \ (3.3)  \times 10^{-10}$~\cite{Bennett:2002jb,PDG2016}  and the SM prediction 
$a_{\mu}^{\rm SM}$ could be an indication of the existence of new
physics beyond the SM. Efforts to improve the experimental estimate at Fermilab
(FNAL)~\cite{Grange:2015fou} and at J-PARC~\cite{Mibe:2010zz,Abe:2019thb} aim to
reduce the experimental uncertainty by a factor of four compared to
the BNL measurement. It is therefore imperative that the SM prediction
is also improved to determine whether the $g-2$ discrepancy is well
established. 

The uncertainty of $a_{\mu}^{\rm SM}$ is entirely dominated by the
hadronic contributions, where the hadronic vacuum polarisation contributions can be
separated into the leading-order (LO) and higher-order contributions. These are calculated utilising dispersion integrals
and the experimentally measured cross section $\sigma^0_{{\rm had},\gamma} (s) \equiv \sigma^0(e^+e^-\rightarrow
\gamma^* \rightarrow {\rm hadrons} + \gamma)$, where the superscript 0 denotes the bare cross section (undressed of
all vacuum polarisation (VP) effects) and the subscript $\gamma$ indicates
the inclusion of effects from final state photon radiation (FSR). At
LO, the dispersion relation reads 
\beq \label{eq:amu}
a_{\mu}^{\rm had,\,LO\,VP} =
\frac{\alpha^2}{3\pi^2}\int^{\infty}_{m_{\pi}^2} \frac{{\rm d}s}{s}
R(s)K(s) \ \ \ \ ; \ \ \ \ R(s) = \frac{\sigma^0_{{\rm had},\gamma} (s)}{\sigma_{\rm pt}(s)}
\equiv \frac{\sigma^0_{{\rm had},\gamma} (s)}{4\pi\alpha^2/3s} \, ,
\eeq
where $R(s)$ denotes the hadronic $R$-ratio and $K(s)$ is a well known kernel function. 

Below $\sim2$GeV, the estimates of $a_{\mu}^{\rm had,\, VP}$ and the corresponding uncertainties are determined from the experimentally measured cross sections of individual hadronic final states, where the hadronic $R$-ratio in this region is predominantly constructed from the sum of the determined cross sections. Above $\sim2$GeV, data for the measured total hadronic $R$-ratio (all hadronic final states) are combined. For nearly all these channels, the available data from numerous different experiments must be analysed, combined and then integrated over according to equation~\eqref{eq:amu} to give a corresponding estimate of the contribution to $a_{\mu}^{\rm had,\, LO\, VP}$. Therefore, the dependence of this calculation on the quality and precision of these measured cross sections is substantial and many experiments have dedicated programmes focused on the accurate measurement of these processes. This document focuses on the effect of these measurements on the recent KNT18 analysis of $a_{\mu}^{\rm had,\, VP}$~\cite{Keshavarzi:2018mgv}. Details of other similar analyses can be found in~\cite{Davier:2017zfy,Jegerlehner:2017gek,Jegerlehner:2017lbd,Benayoun:2015gxa,Colangelo:2018mtw}.

\section{Experimental measurements of $e^+e^-\rightarrow {\rm hadrons}$}

Experimental measurements of the cross sections of exclusive hadronic final states are obtained via two approaches. The first is the standard direct energy scan approach, where data is collected at fixed centre of mass (C.M.) energy intervals. The second is achieved through radiative return, where the differential cross section is measured as a function of the invariant mass of the hadronic final state, $\sqrt{s} = M_{\rm had}$. The cross section $\sigma_{\rm had} \equiv \sigma(e^+e^-\rightarrow \ {\rm hadrons})$ is then determined, for example, according to~\cite{Binner:1999bt} using the relation
\beq \label{pipidiffxSec}
s\frac{{\rm d}\sigma\big({\rm had}+\gamma\big)}{dM_{\rm had}^2} = \sigma_{\rm had}(M_{\rm had}^2)H(M_{\rm had}^2,s) \ ,
\eeq
where $H$ is the radiator function describing the emission of photons in the initial state~\cite{Rodrigo:2001kf,Czyz:2002np,Czyz:2003ue,Czyz:2004rj}.

\subsection{Direct energy scan experiments}

\subsubsection{CMD-3}
The CMD-3 detector~\cite{Khazin:2008zz} is the first of two direct energy scan experiments at the $e^+e^-$ collider VEPP-2000~\cite{Khazin:2010zz}. The VEPP-2000 machine has a C.M. energy range of $0.3 \leq \sqrt{s} \leq 2$ GeV, with a design luminosity of $L = 10^{32}cm^{-2}s^{-1}$ at $\sqrt{s} = 2$ GeV. The CMD-3 experimental programme has already published cross section measurements for many final states (see e.g.~\cite{Akhmetshin:2018mqd,Solodov:2017pyu,CMD-3_PhiPsi19}). Of these, major improvements have been seen in the measurements of the $K\bar{K}$ cross sections, with both the $K^+K^-$~\cite{Kozyrev:2017agm} and $K^0_S K^0_L$~\cite{Kozyrev:2016raz} analyses yielding very precise results of the narrow $\phi$ resonance that dominates in both these channels. In the $K^+K^-$ channel in particular, these new data replace those previously measured by CMD-2~\cite{Akhmetshin:2008gz}, which are currently awaiting reanalysis as they suffer from an overestimation of the trigger efficiency for slow kaons~\cite{Kozyrev:2017agm,CMD-2trigger}. The cross section values of these new CMD-3 data are higher than all other existing data in this channel~\cite{Kozyrev:2017agm}, leading to significant new data tensions in the overall combination of all available $K^+K^-$ data (see Section~\ref{sec:DataTensions}). Notably, the CMD-3 experiment has also recently released data for the $3\pi^+3\pi^-\pi^0$ final state, which had not previously been measured~\cite{CMD-3:2019ufp}.
% This provides an almost negligible contribution to $a_{\mu}^{\rm had,\, VP}$, with the production threshold for seven pions being very close to the $\sim2$ GeV transition point between the sum exclusive final states and the inclusive hadronic $R$-ratio data. It is, however, reassuring that this previously omitted contribution is small enough that its absence in earlier determinations of $a_{\mu}^{\rm had,\, VP}$ is of no consequence.
Of particular importance to future determinations of $a_{\mu}^{\rm had,\, VP}$ is the announced new measurement of the $\pi^+\pi^-$ cross section by CMD-3, which is currently undergoing an extensive analysis~\cite{CMD-3_PhiPsi19}.\footnote{The $\pi^+\pi^-$ channel accounts for over 70\% of the total value of $a_{\mu}^{\rm had, \, LO \, VP}$, due to the large $\rho$ resonance structure in the low energy region below 1 GeV that is highly weighted by $K(s)$ in equation~\eqref{eq:amu}. Consequently, it also dominates the overall uncertainty of the hadronic vacuum polarisation contributions, resulting in CMD-3 (and other experiments) re-measuring this final state in an attempt to more precisely determine $a_\mu^{\rm had, \, VP}$.} With high-precision in mind, this measurement aims to be the most precise in terms of statistical precision of all the current data sets being combined in the $\pi^+\pi^-$ channel and to also achieve a systematic uncertainty budget of $\sim0.4-0.5\%$, compared to the $\sim0.6-0.8\%$ achieved by CMD-2~\cite{Akhmetshin:2006wh,Akhmetshin:2006bx,Aulchenko:2006na}.

\subsubsection{SND}

The SND experiment~\cite{Achasov:2009zza} is the second general purpose detector at VEPP-2000~\cite{Khazin:2010zz}. It also collected data at the VEPP-2M collider~\cite{Aulchenko:1975dc} that predated this, where data for exclusive hadronic final states were collected between 1996-2000 in the energy range $0.4\leq\sqrt{s}\leq1.4$ GeV. This was then extended to $0.3\leq\sqrt{s}\leq2.0$ GeV as part of the upgrade to the VEPP-2000 machine. In recent years, SND have released new data for several hadronic modes~\cite{Kupich:2017luv,Logashenko:2016xhy}, notably the $\pi^0\gamma$~\cite{Achasov:2016bfr,Achasov:2018ujw} cross section and the $K^+K^-$ cross section above the $\phi$ resonance~\cite{Achasov:2016lbc}. The SND experiment is also currently analysing a new measurement of the $\pi^+\pi^-$ cross section, having collected an integrated luminosity of $5{\rm pb}^{-1}$ of data for this important final state~\cite{SND_PhiPsi19}. The systematic uncertainties of this measurement are predicted to be in the range of $0.8-0.9\%$.

\subsubsection{KEDR}

The KEDR detector~\cite{KEDR} at the VEPP-4M $e^+e^-$ collider~\cite{VEPP-4M} is an experimental facility dedicated to the measurement of the full multi-hadron cross section, or total hadronic $R$-ratio. It has already published measurements of $R(s)$ at 22 C.M. energies between $1.84\leq\sqrt{s}\leq3.72$ GeV, with total uncertainties ranging from $3.9\%$ (2.4\% systematic uncertainties) at lower energies to 2.6\% ($\sim1.9\%$ systematic uncertainties) above the $J/\psi$ resonance~\cite{KEDR_PhiPsi19}. The agreement between these data and pQCD in this energy range is much improved compared to the previous measurements of the $R$-ratio by BES/BES-II in this region\cite{Bai:1999pk,Bai:2001ct,Ablikim:2006aj,Ablikim:2006mb,Ablikim:2009ad}. The KEDR experiment also plans to complete two scans of $R(s)$ from $4.56\leq\sqrt{s}\leq6.96$ GeV by the end of 2019~\cite{KEDR_PhiPsi19}.

\subsection{Radiative return experiments}

\subsubsection{BABAR}

The BABAR detector~\cite{Aubert:2001tu} resides at the PEP-II $e^+e^-$ storage ring at SLAC~\cite{PEP-II}, which operates predominantly at the C.M. energy $\sqrt{s} = 10.6$ GeV. The experiment detects large-angle photons with energies $E^*_{\gamma} > 3$ GeV, which defines the C.M. energy $\sqrt{s'}$ of the leptonic or hadronic final state determined via radiative return. This allows for precise cross sections measurements from production threshold up to 3-5 GeV~\cite{Michel_BABAR}. 

The experimental programme at BABAR dedicated to low-energy hadronic cross sections has measured an almost complete set of exclusive hadronic channels below 2 GeV, missing only the $\pi^+\pi^-\pi^0$, $\pi^+\pi^-4\pi^0$ and $\geq7\pi$ modes. Arguably its most impressive measurement is that of the $\pi^+\pi^-(\gamma)$ cross section from threshold to $\sqrt{s'} \leq 3$ GeV, which is statistically the most precise of all measured $\pi^+\pi^-$ cross sections and has a systematic uncertainty of only 0.5\% in the region of the all-important $\rho$ resonance~\cite{Aubert:2009ad,Lees:2012cj}. BABAR have also announced a forthcoming release of a new measurement of the $\pi^+\pi^-(\gamma)$ cross section which should have twice the number of the statistics of the data published in~\cite{Lees:2012cj} and have even better control of the systematic uncertainties~\cite{Michel_BABAR}.

With their broad experimental programme, BABAR measurements have also greatly improved the determination of many other channels. A new measurement of the $\pi^+\pi^-\pi^0\pi^0$ channel~\cite{TheBABAR:2017vzo} has provided the only new data in this channel since 2003. The $K^+K^-$ channel now includes a precise and finely-binned measurement, supplemented with full statistical and systematic covariance matrices~\cite{Lees:2013gzt}. The neutral final state $K^0_S K^0_L\pi^0$ has also been measured, completing all modes that contribute to the $KK\pi$ final state. In addition, BABAR have also completed all modes that contribute to the $KK\pi\pi$ channel~\cite{TheBABAR:2017vgl}. Finally, a very recent measurement of the $\pi^+\pi^-3\pi^0$ cross section is the first published data of this final state to be included in the overall compilation~\cite{Lees:2018dnv}.

\subsubsection{KLOE/KLOE-2}

DA$\Phi$NE~\cite{DAFNE} is a high luminosity $e^+e^-$ collider that operates predominantly at the centre of mass energy equal to the $\phi$ meson mass, $\sqrt{s} = m_{\phi} = 1.0194 \text{ GeV} $~\cite{PDG2016}. The KLOE detector has been used to obtain three precise measurements of the cross section $\allowbreak\sigma\big(e^+e^-\allowbreak\rightarrow\pi^+\pi^-\gamma(\gamma)\big)$ in 2008~\cite{Ambrosino:2008aa,KLOE08-KLOEnote}, 2010~\cite{Ambrosino:2010bv,KLOE10-KLOEnote} and 2012~\cite{Babusci:2012rp,KLOE12-KLOEnote}.\footnote{The KLOE collaboration also made a measurement of $\sigma\big(e^+e^-\rightarrow\pi^+\pi^-\gamma(\gamma)\big)$ in 2005~\cite{Aloisio:2004bu}. However, this is now considered to be superseded by the 2008 measurement, as discussed in~\cite{Ambrosino:2008aa}.} Each of these measurements results in two-pion contribution to the anomalous magnetic moment of the muon of~\cite{Anastasi:2017eio}
\begin{align}
a_{\mu}^{\pi^+\pi^-}({\rm KLOE08}, \ 0.35< s <0.95 \text{ GeV}^2) & = (386.6 \pm 0.4_{\rm stat} \pm 3.3_{\rm sys}) \times 10^{-10} , \nonumber
\\
\
a_{\mu}^{\pi^+\pi^-}({\rm KLOE10}, \ 0.10 < s <0.85 \text{ GeV}^2) & = (477.9 \pm 2.0_{\rm stat} \pm 6.7_{\rm sys}) \times 10^{-10} , \nonumber
\\
\
a_{\mu}^{\pi^+\pi^-}({\rm KLOE12}, \ 0.35< s <0.95 \text{ GeV}^2) & = (385.1 \pm 1.2_{\rm stat} \pm 2.3_{\rm sys}) \times 10^{-10} .
\end{align}
The simultaneous use of the KLOE measurements required a detailed analysis to attain the correct combination of the three, which have a non-trivial influence on $a_{\mu}^{\pi^+\pi^-}$ and provide an important comparison with other experimental measurements of $\sigma_{\pi\pi}$. The KLOE measurements of $\sigma_{\pi\pi(\gamma)}$ are, in part, highly correlated, necessitating the construction of full statistical and systematic covariance matrices to be used in any combination of these data. The construction of these matrices and the combination of the three measurements was achieved in~\cite{Anastasi:2017eio}, which resulted in a single vector for the two-pion cross section $\sigma_{\pi\pi(\gamma)}$, along with a corresponding covariance matrix. This combination of the KLOE cross section data resulted in an estimate of the two-pion contribution to the anomalous magnetic moment of the muon of
\beq \label{KLOEcombination}
a_{\mu}^{\pi^+\pi^-}({\rm KLOE \ combination}, \ 0.10< s <0.95 \text{ GeV}^2) = (489.8 \pm 1.7_{\rm stat} \pm 4.8_{\rm sys} ) \times 10^{-10} .
\eeq
\subsubsection{BESIII}

The BESIII detector~\cite{Ablikim:2009aa} is a general purpose detector at the BEPCII $e^+e^-$ collider~\cite{BEPCII}, which operates at C.M. energies between $2.0\leq\sqrt{s}\leq4.6$ GeV and has a design luminosity of $L = 10.0^{33}cm^{-2}s^{-1}$ at the $\psi(3770)$ resonance~\cite{BESIII_Redmer}. The BESIII experiment have published a measurement of the $e^+e^-\rightarrow\pi^+\pi^-$ cross section focused solely on the $\rho$ resonance contribution between $0.6\leq\sqrt{s}\leq0.9$ GeV~\cite{Ablikim:2015orh}. With a total uncertainty of $\sim0.9\%$ and the evident disagreement between the BABAR and KLOE cross sections, this measurement provides an interesting comparison to the existing data. BESIII have also announced future releases of the $\pi^+\pi^-\pi^0$, $\pi^+\pi^-\pi^0\pi^0$, $\omega\pi^0$ and $\pi^+\pi^-3\pi^0$ cross sections~\cite{BESIII_TGm2}, along with measurements of the total hadronic $R$-ratio~\cite{BESIII_PhiPsi19}.

\section{Results for $a_{\mu}^{\rm had,\, VP}$ from KNT18}

\begin{table}[!t]
\centering
%\vspace{-1cm}
%\hspace{2.0cm}
\scalebox{0.9}{
 {\renewcommand{\arraystretch}{0.9}
 \begin{tabular}{|l|c|c|}
\hline																						
{\bf Channel}	&	{\bf Energy range (GeV)}							&	$a_{\mu}^{\rm had, \, LO \, VP} \times 10^{10}$						\\
\hline			
$\pi^+\pi^-$	&	$	0.305	\leq	\sqrt{s}	\leq	1.937	$	&	$	502.97	\pm	1.97\hphantom{0}	$		\\								$\pi^+\pi^-\pi^0$	&	$	0.660	\leq	\sqrt{s}	\leq	1.937	$	&	$	47.79	\pm	0.89	$		\\
$\pi^+\pi^-\pi^+\pi^-$	&	$	0.613	\leq	\sqrt{s}	\leq	1.937	$	&	$	14.87	\pm	0.20	$	\\
$\pi^+\pi^-\pi^0\pi^0$	&	$	0.850	\leq	\sqrt{s}	\leq	1.937	$	&	$	19.39	\pm	0.78	$		\\
$K^+K^-$	&	$	0.988	\leq	\sqrt{s}	\leq	1.937	$	&	$	23.03	\pm	0.22	$\\
$K^0_S K^0_L$	&	$	1.004	\leq	\sqrt{s}	\leq	1.937	$	&	$	13.04	\pm	0.19	$		\\
$KK\pi$	&	$	1.260	\leq	\sqrt{s}	\leq	1.937	$	&	$	\hphantom{0}2.71	\pm	0.12	$		\\
$KK2\pi$	&	$	1.350	\leq	\sqrt{s}	\leq	1.937	$	&	$	\hphantom{0}1.93	\pm	0.08	$			\\
Inclusive channel	&	$	1.937	\leq	\sqrt{s}	\leq	11.200	$	&	$	43.67	\pm	0.67	$		\\
\hline																						\end{tabular} 
} 
}\caption{Contributions to $a_{\mu}^{\rm had, \, LO \, VP}$~\cite{Keshavarzi:2018mgv}.}\label{tab:amuhadexc}
\end{table}
 \begin{figure}[!t]
\centering
  \subfloat[$\sigma^{0}(e^+e^-\rightarrow\pi^+\pi^-)$]{%
    \includegraphics[width= 0.33\textwidth]{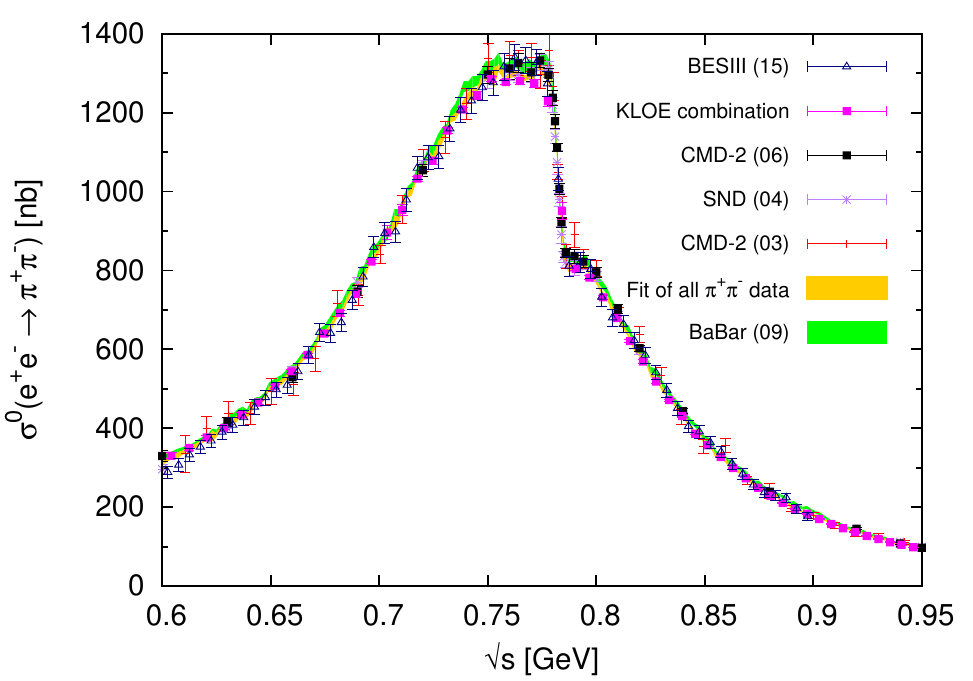}}\hfill
  \subfloat[$\sigma^{0}(e^+e^-\rightarrow\pi^+\pi^-\pi^0)$]{%
    \includegraphics[width= 0.33\textwidth]{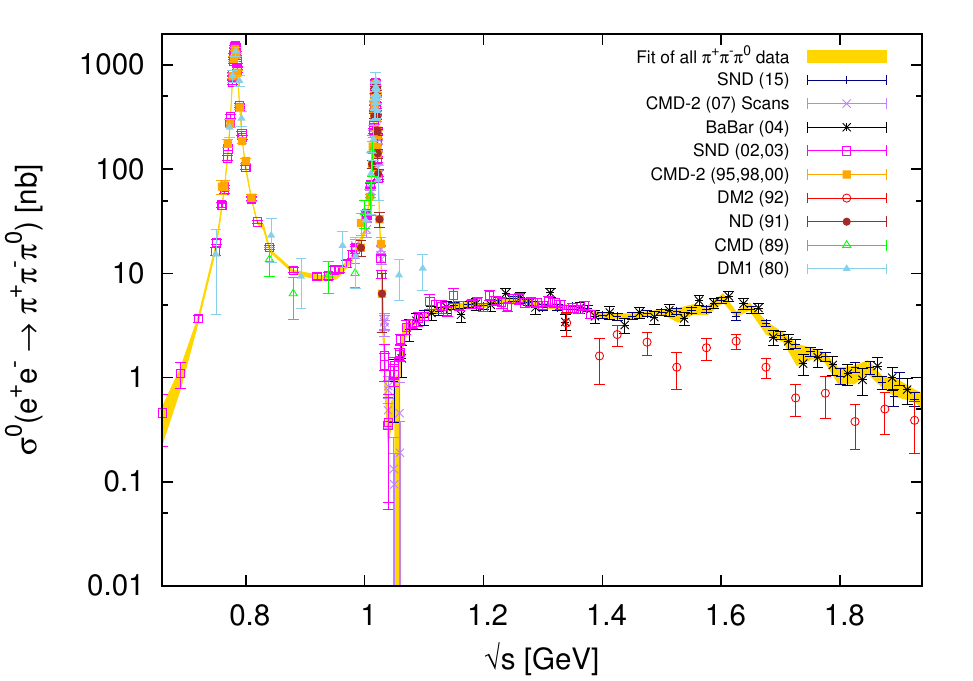}}\hfill
  \subfloat[$\sigma^{0}(e^+e^-\rightarrow\pi^+\pi^-\pi^+\pi^-)$]{%
    \includegraphics[width= 0.33\textwidth]{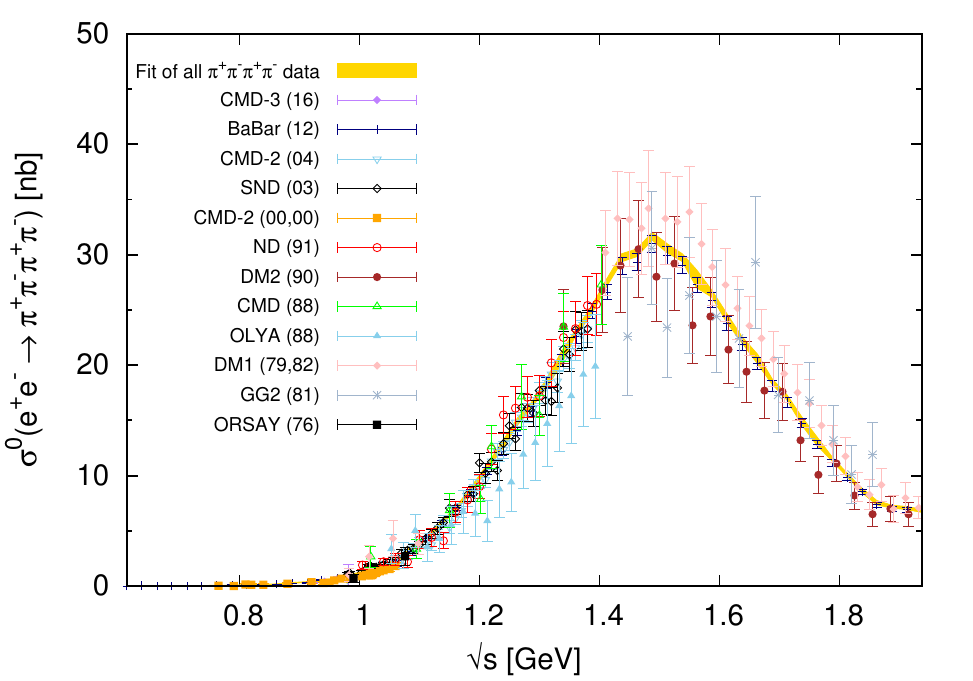}}\hfill
  \subfloat[$\sigma^{0}(e^+e^-\rightarrow\pi^+\pi^-\pi^0\pi^0)$]{%
    \includegraphics[width= 0.33\textwidth]{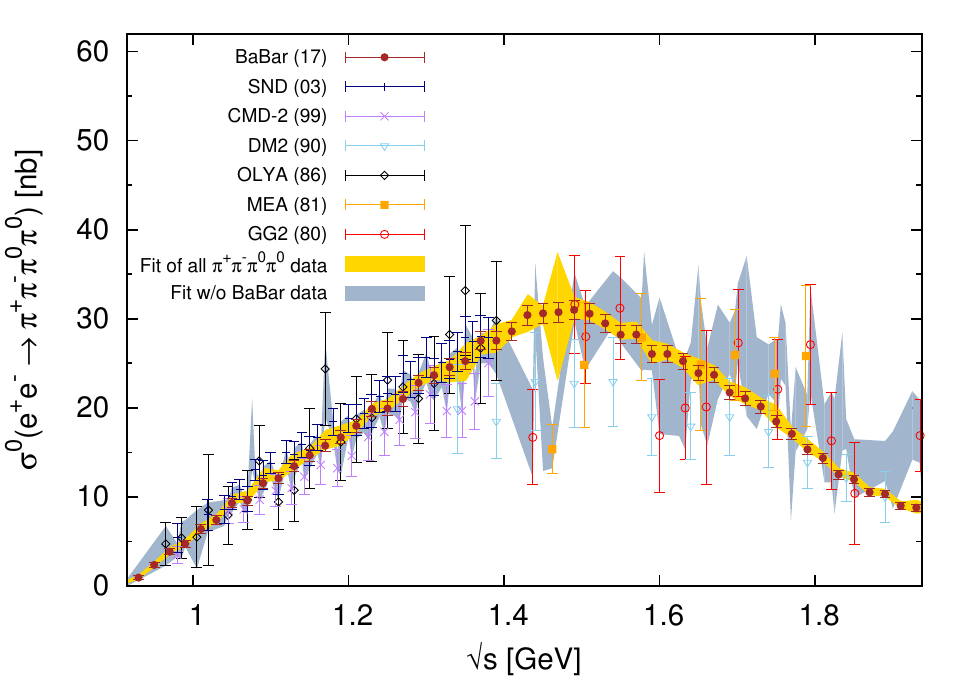}}\hfill
  \subfloat[$\sigma^{0}(e^+e^-\rightarrow K^+K^-)$]{%
    \includegraphics[width= 0.33\textwidth]{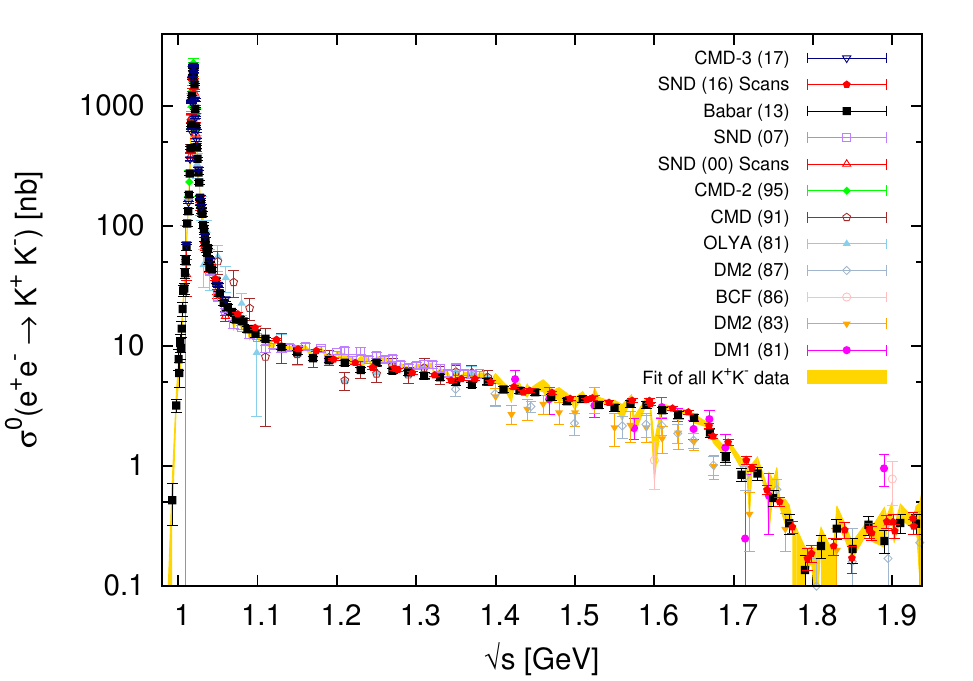}}\hfill
  \subfloat[$\sigma^{0}(e^+e^-\rightarrow K^0_S K^0_L)$]{%
    \includegraphics[width= 0.33\textwidth]{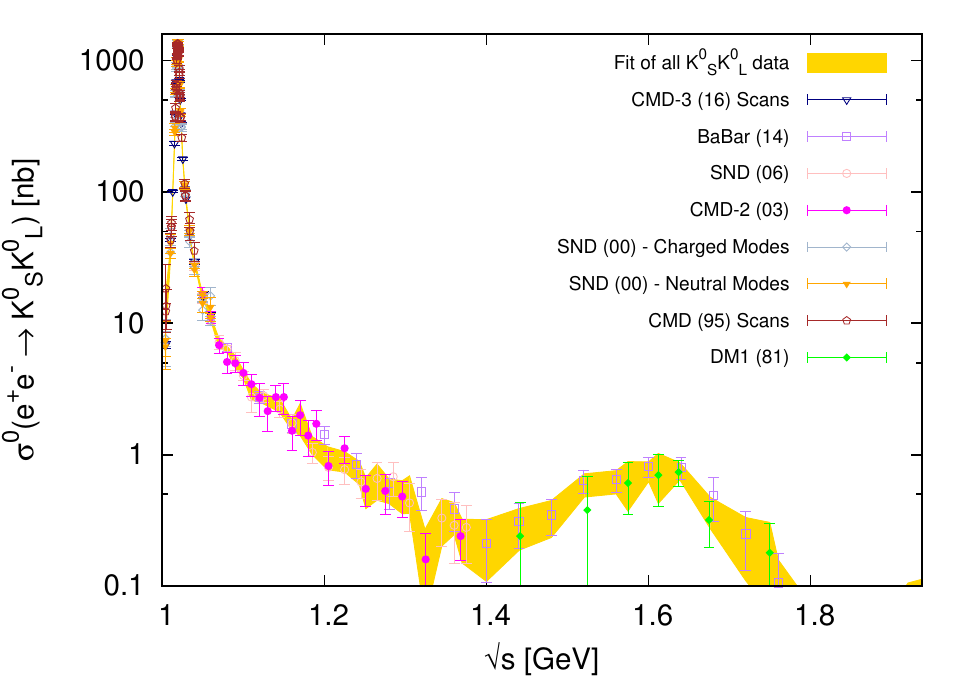}}\hfill
  \subfloat[$\sigma^{0}(e^+e^-\rightarrow KK\pi)$]{%
    \includegraphics[width= 0.33\textwidth]{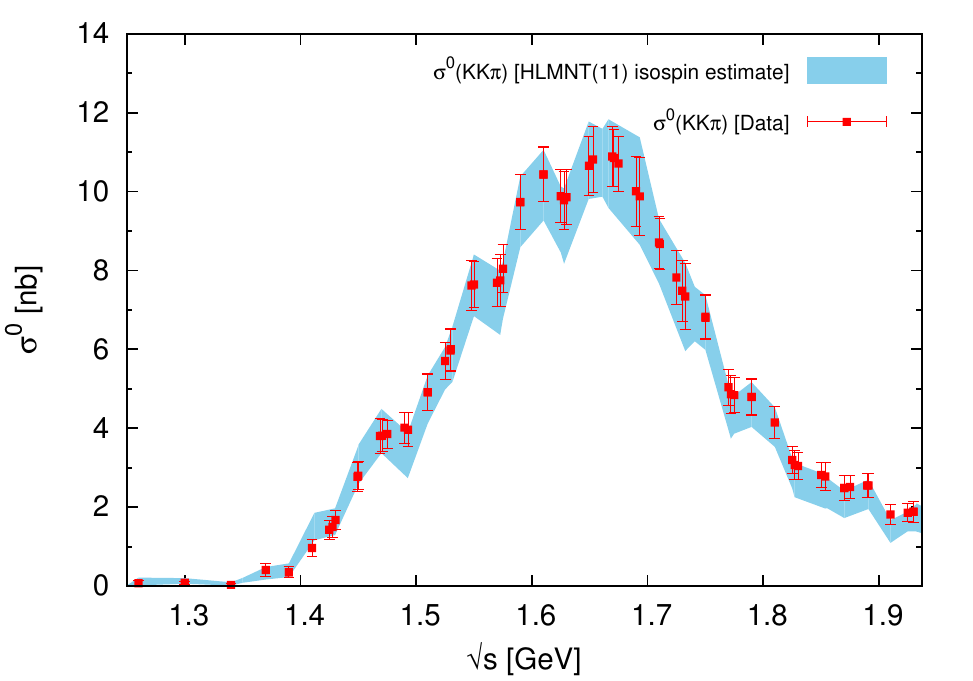}}\hfill
  \subfloat[$\sigma^{0}(e^+e^-\rightarrow KK\pi\pi)$]{%
    \includegraphics[width= 0.33\textwidth]{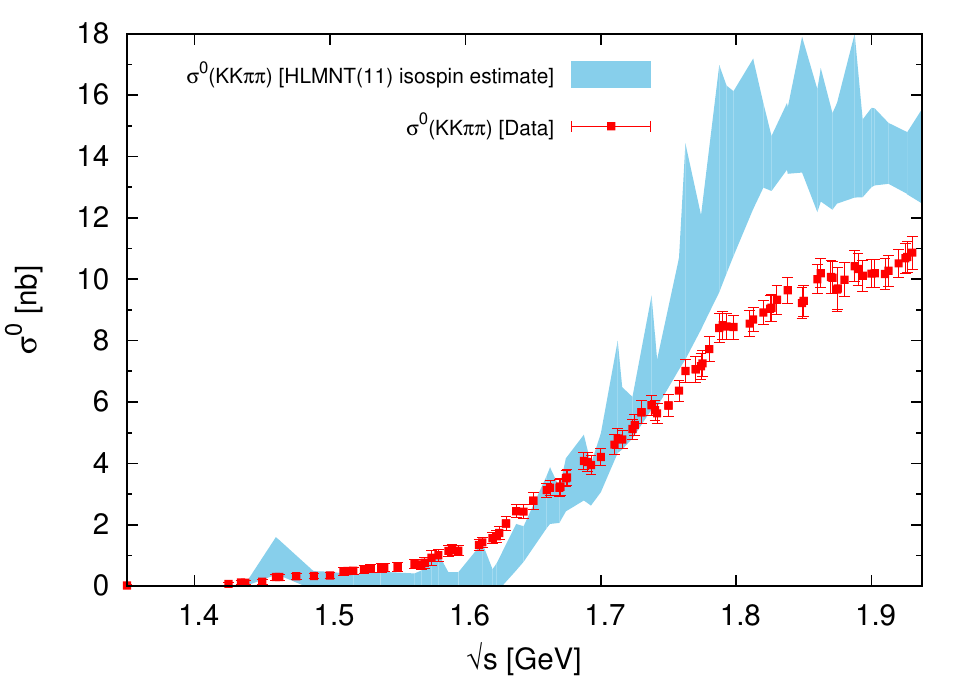}}\hfill
  \subfloat[Inclusive data]{%
    \includegraphics[width= 0.33\textwidth]{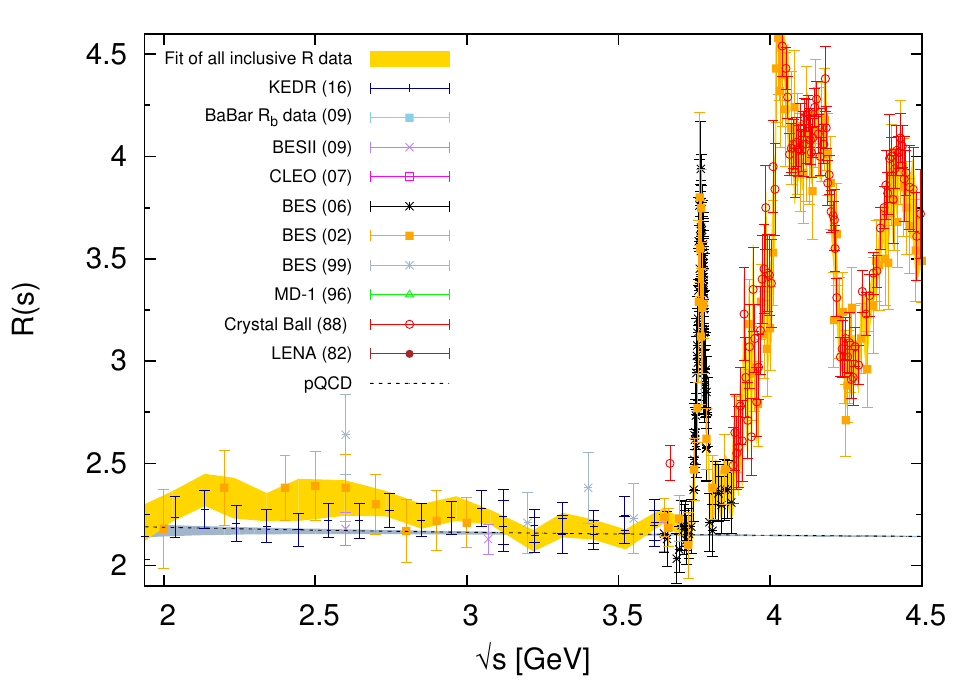}}\hfill
  \caption{The resulting cross sections of the leading and major sub-leading hadronic final states~\cite{Keshavarzi:2018mgv}.}\label{fig:excxSec}
\end{figure}
The KNT18 analysis~\cite{Keshavarzi:2018mgv} is a complete re-evaluation of the hadronic vacuum polarisation contributions, $a_{\mu}^{\rm had, \ VP}$. The results from this work for contributions to $a_{\mu}^{\rm had, \, LO \, VP}$ and cross sections from the major channels are given in Table~\ref{tab:amuhadexc} and Figure~\ref{fig:excxSec}, respectively. In the  $\pi^+\pi^-$ channel, the radiative return measurements from BABAR, KLOE and BESIII in the $\rho$ region have greatly improved the estimate of this final state. The cross section in the $\rho$ region is displayed in plot (a) of Figure~\ref{fig:excxSec}. For all displayed channels, the data combinations include new measurements which, coupled with updates in the KNT data combination routine~\cite{Keshavarzi:2018mgv}, have improved the estimates of $a_{\mu}^{\rm had, \, LO \, VP}$ from these final states. The uncertainty contribution from $\pi^+\pi^-\pi^0\pi^0$ is still relatively large in comparison with its contribution to $a_{\mu}^{\rm had, \, LO \, VP}$ and requires better new data. Plot (g) of Figure~\ref{fig:excxSec} demonstrates good agreement between the previously used estimate from isospin relations and the data-based approach in the $KK\pi$ final state. Examining plot (h) of Figure~\ref{fig:excxSec}, it is evident that the isospin relations provided a poor estimate of the $KK\pi\pi$ final state. The inclusive hadronic $R$-ratio compilation is shown in plot (i) of Figure~\ref{fig:excxSec}, which demonstrates that the inclusive data combination is much improved. With the new KEDR data, the differences between the inclusive data and pQCD are not as large as previously and, hence, the contributions in the entire inclusive data region are now estimated using the inclusive data alone. 

\subsection{Data tensions in the KNT18 analysis}\label{sec:DataTensions in the KNT}

\begin{figure}[!t] 
  \centering
    {\includegraphics[width=0.5\textwidth]{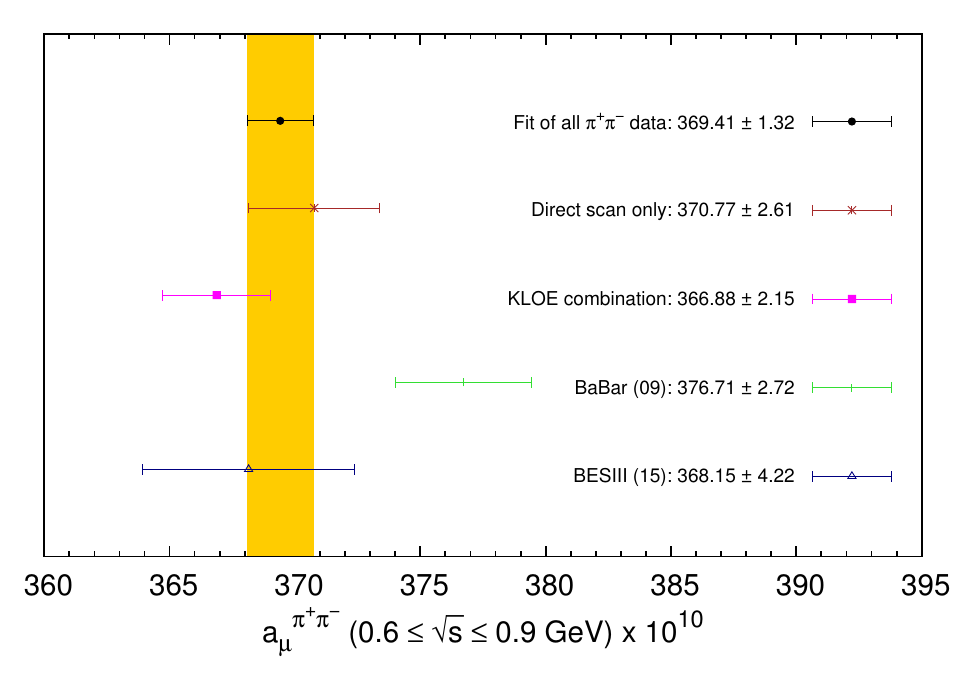}
     \caption{\small The comparison of the integration of the individual radiative return measurements and the combination of direct scan $\pi^+\pi^-$ measurements between $0.6 \leq \sqrt{s} \leq 0.9$ GeV~\cite{Keshavarzi:2018mgv}.} \label{fig:RadRetCompare}}
\end{figure} 

 \begin{figure}[!t]
\centering
  \subfloat[$\pi^+\pi^-$]{%
    \includegraphics[width= 0.535\textwidth]{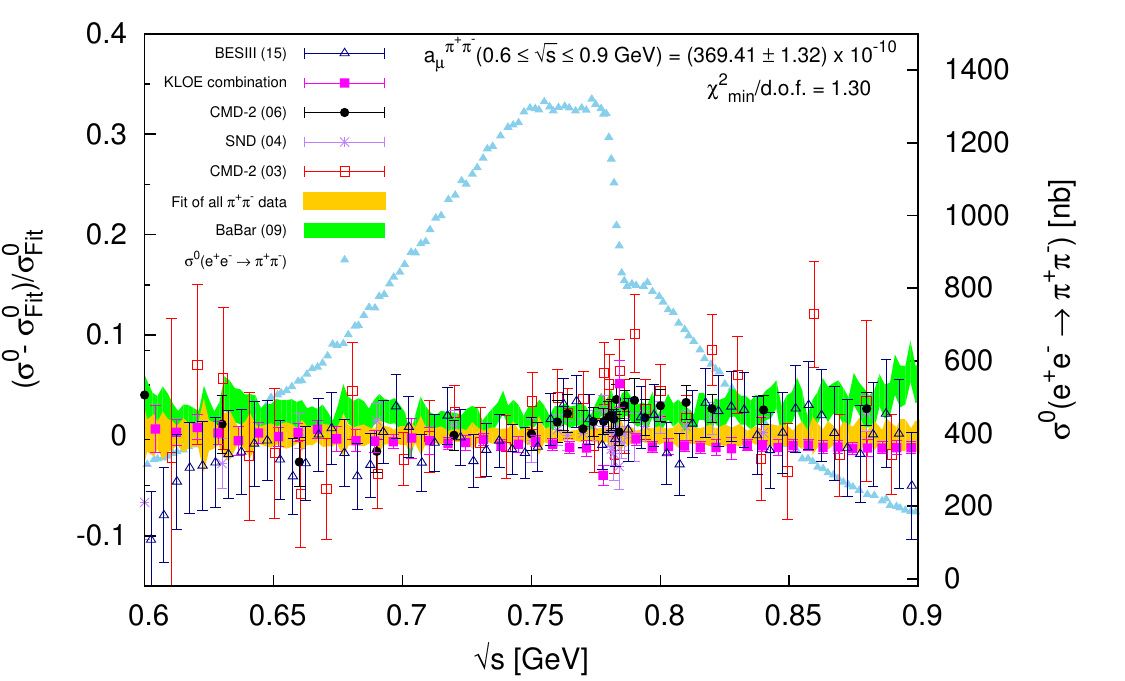}\label{fig:RadRetFit}}\hfill
  \subfloat[$K^+K^-$]{%
    \includegraphics[width= 0.465\textwidth]{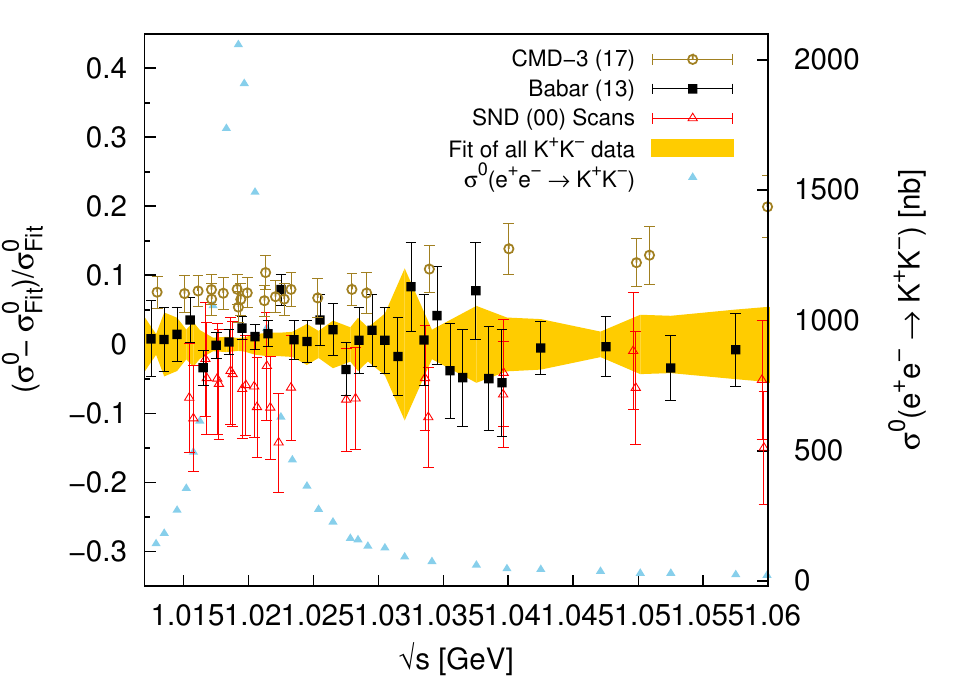}\label{fig:K+K-diff}}\hfill
  \caption{Relative difference plots of data in the $\pi^+\pi^-$ channel on the $\rho$ resonance and in the $K^+K^-$ channel on the $\phi$ resonance, against the fit all of all data for the respective channel. The width of the coloured bands represent the propagation of the statistical and systematic uncertainties added in quadrature~\cite{Keshavarzi:2018mgv}.}
\end{figure}

%
%\begin{figure}[!t] 
%  \centering
%    {\includegraphics[width=0.6\textwidth]{Plots/Section3/Data/2pi/Ch10-KNT2017_Diff-RhoInt-eps-converted-to.pdf}
%     \caption{The relative difference of the radiative return and most relevant direct scan data sets contributing to $a_{\mu}^{\pi^+\pi^-}$ and the combination of all data, plotted in the $\rho$ region. The width of the coloured bands represent the propagation of the statistical and systematic uncertainties added in quadrature.}     \label{fig:RadRetFit}}
%\end{figure} 
%\begin{figure}[!t] 
%\centering
%\includegraphics[width=0.75\textwidth]{Plots/Section3/Data/KK/Ch02_Diff_phi-eps-converted-to.pdf}
%\caption{ The relative difference of the dominant data in the $K^+K^-$ channel measured in the $\phi$ resonance region and the fit of all data. For comparison, the individual sets have been normalised against the fit. The yellow band represents the full data combination which incorporates all correlated statistical and systematic uncertainties. However, the width of the yellow band simply displays the square root of the diagonal elements of the total output covariance matrix of the fit.} \label{fig:K+K-diff}
%\end{figure}

In the $\pi^+\pi^-$ channel, the BABAR data are noticeably higher than the average, causing an increase to the two-pion contribution to $a_{\mu}^{\rm SM}$. This is evident from Figure~\ref{fig:RadRetCompare}, which compares the estimates of $a_{\mu}^{\pi^+\pi^-}$ from the full data combination, the radiative return measurements and all other measurements (direct energy scan) in the dominant $\rho$ region. Notably, the deviation between the estimates from KLOE combination and the BABAR data in this range is $\sim 2.8\sigma$. With the highly correlated nature of the KLOE combination now having a dominating influence overall in the KNT18 analysis, a large disagreement is also noted between the full $\pi^+\pi^-$ data combination and the integral of the BABAR data alone.\footnote{This effect is more prominent when considering the evaluation of $a_{\mu}^{\pi^+\pi^-}$ from the BABAR data alone over the full available energy range. This results in an estimate of $a_{\mu}^{\pi^+\pi^-}(\text{BaBar data only})  =  513.2 \pm 3.8$ compared to the result given in Table~\ref{tab:amuhadexc}. It should be noted that similar differences are observed between the integral of the BABAR data alone and full evaluations of $a_{\mu}^{\pi^+\pi^-}$ from other recent analyses~\cite{Davier:2017zfy,Jegerlehner:2017gek,Jegerlehner:2017lbd,Benayoun:2015gxa,Colangelo:2018mtw}.} This is made more apparent when considering Figure~\ref{fig:RadRetFit}. From Figure~\ref{fig:RadRetCompare}, is clear that the full $\pi^+\pi^-$ data combination agrees well with the new BESIII data, the KLOE data and the combination of the remaining direct scan data. Interestingly, however, the BESIII data is in better agreement with the BABAR data at the peak of the resonance where the cross section is largest, slightly alleviating the disagreement between the full $\pi^+\pi^-$ data combination and the BABAR data. The tension between data sets is, however, reflected and accounted for in the local $\chi^2$ error inflation, which results in an $\sim15\%$ increase in the uncertainty of $a_{\mu}^{\pi^+\pi^-}$~\cite{Keshavarzi:2018mgv}. 

This estimate of $a_{\mu}^{K^+K^-}$ exhibits an increase of the mean value of more than 1$\sigma$ attributed to the inclusion of the new BABAR and CMD-3 data. This can be seen in Figure~\ref{fig:K+K-diff}. Previously, the data combination in the $\phi$ resonance region for this channel was dominated by the SND scans~\cite{Achasov:2000am} visible in Figure~\ref{fig:K+K-diff} and the now omitted CMD-2 scans~\cite{Akhmetshin:2008gz}, which were in good agreement. The BABAR data~\cite{Lees:2013gzt}, which due to their precision and correlated uncertainties now dominate the $K^+K^-$ data combination, are higher in this region than both the SND and CMD-2 data. The most recent CMD-3 data are higher still~\cite{Kozyrev:2017agm}. The reanalysis of the CMD-2 data will prove crucial in resolving the current differences in this channel and, should they agree further with the BABAR and CMD-3 data, would result in a further increase of the estimate from this channel. Overall, the uncertainty has drastically improved, with much of the change being due to a finer clustering over the $\phi$ resonance after the inclusion of the new high statistics BABAR data. However, the disagreement between the data seen in Figure~\ref{fig:K+K-diff} is accounted for in the local error inflation which provides an increase to the uncertainty of $a_{\mu}^{K^+K^-}$ of $\sim20\%$~\cite{Keshavarzi:2018mgv}. 

\subsection{Total results for $a_{\mu}^{\rm had,\,LO\,VP}$ and $a_{\mu}^{\rm had,\,NLO\,VP}$}

% \begin{figure}[!t]
%\centering
%    \includegraphics[width= 0.6\textwidth]{Plots/Section3/Full/pie_amu.png}
%  \caption{Pie charts showing the fractional contributions to the total mean value and (error)$^2$ of both $a_{\mu}^{\rm had, \, LO \, VP}$ from various energy intervals.}  \label{piechart}
%\end{figure} 
 \begin{figure}[!t]
\centering
  \subfloat[The hadronic $R$ ratio]{%
    \includegraphics[width= 0.49\textwidth]{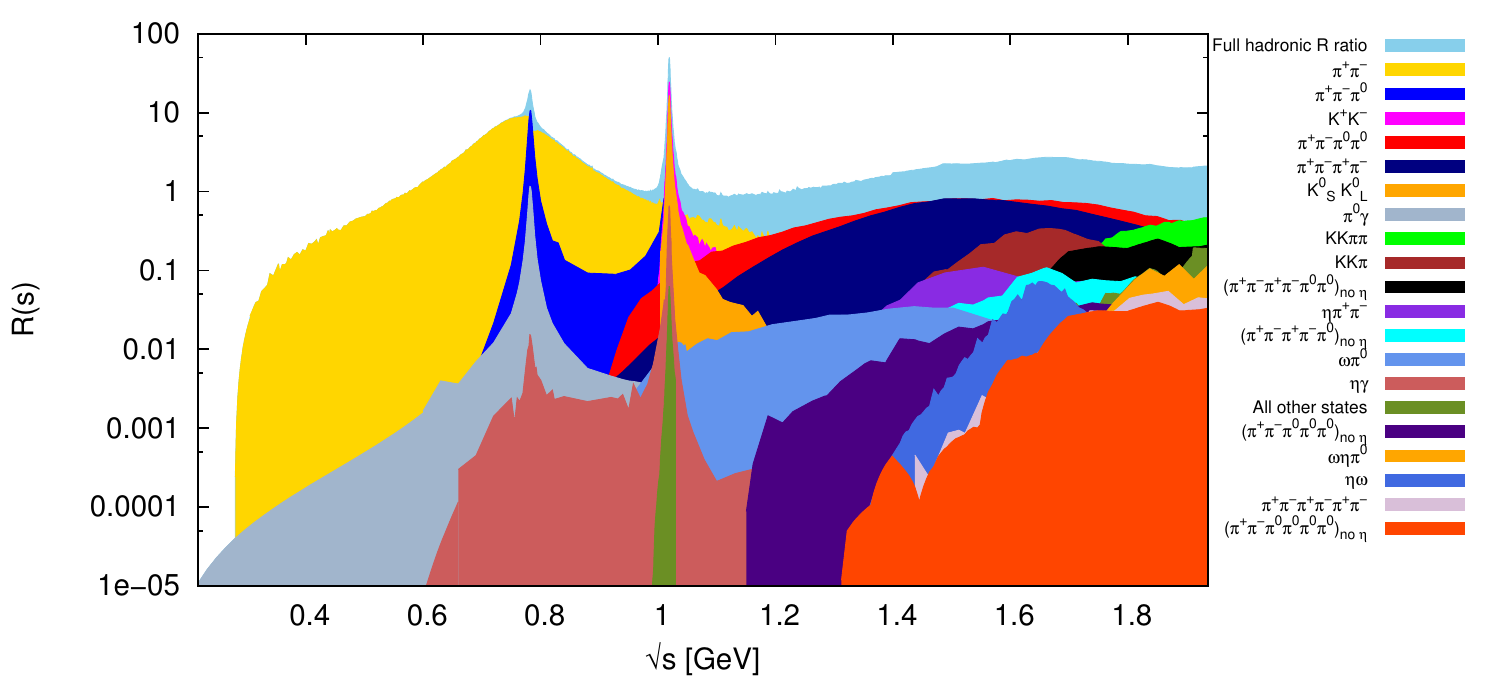}}\hfill
    \vspace{0.5cm}
  \subfloat[The uncertainty of the hadronic $R$ ratio]{%
    \includegraphics[width= 0.49\textwidth]{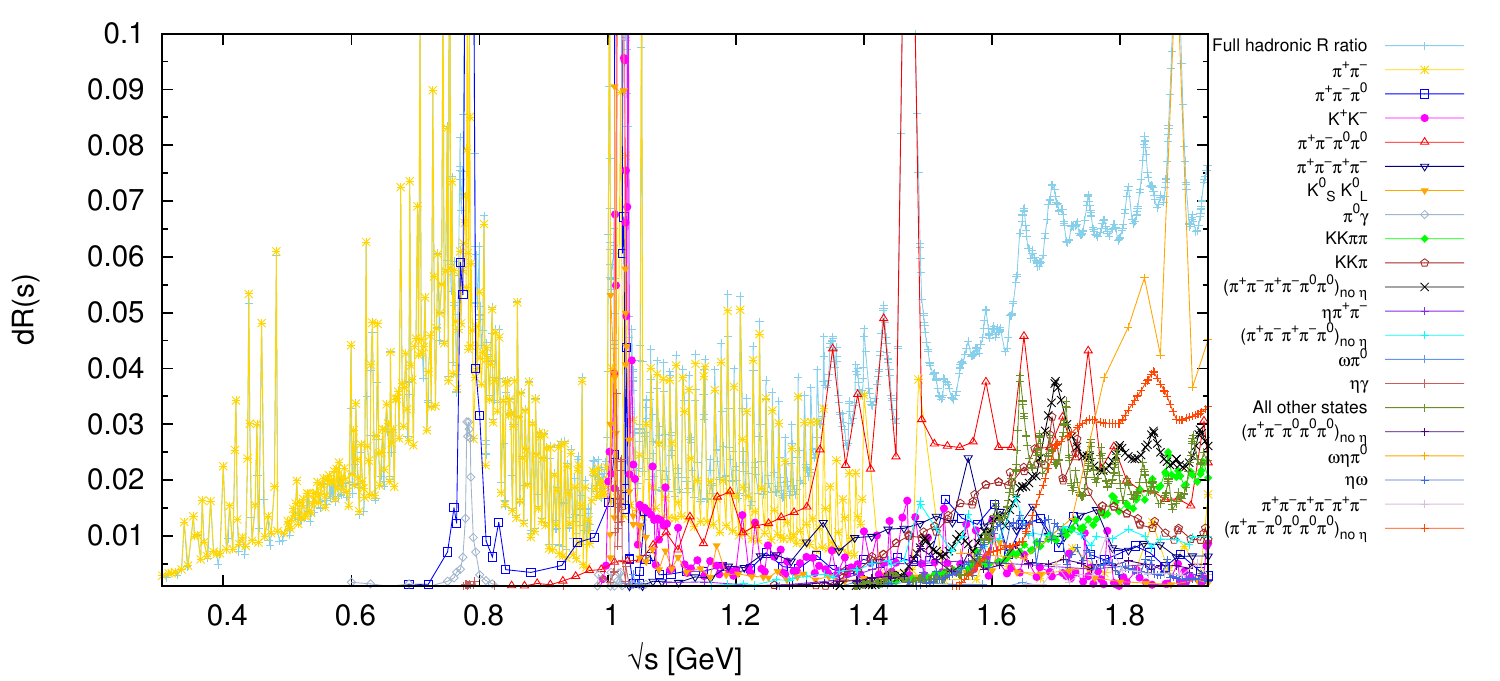}}\hfill
\vspace{-0.5cm}
  \caption{Contributions to the total hadronic $R$ ratio from the different final states (left panel) and their uncertainties (right panel) below 1.937 GeV. The full $R$ ratio and its uncertainty is shown in light blue in each plot, respectively. Each final state is included as a new layer on top in decreasing order of the size of its contribution to $a_{\mu}^{\rm had, \, LO \, VP}$~\cite{Keshavarzi:2018mgv}.}  \label{hadxSec}
\end{figure} 
From the sum of all hadronic contributions shown in Figure~\ref{hadxSec}, the estimate for $a_{\mu}^{\rm had, \, LO \, VP}$ from this analysis is~\cite{Keshavarzi:2018mgv}
\beq\label{LOHVP_KNT18}
a_{\mu}^{\rm had, \, LO \, VP} = (693.26  \pm 1.19_{\rm stat} \pm 2.01_{\rm sys} \pm 0.22_{\rm vp} \pm 0.71_{\rm fsr}) \times 10^{-10} =  (693.26 \pm 2.46_{\rm tot}) \times 10^{-10} \ , 
\eeq
where the uncertainties include all available correlations and local $\chi^2_{\rm min}/{\rm d.o.f.}$ inflation.
Using the same data compilation as for the calculation of $a_{\mu}^{\rm had, \, LO \, VP}$, the next-to-leading order (NLO) contribution to $a_{\mu}^{\rm had, VP}$ is determined to be $a_{\mu}^{\rm had, NLOVP} =  (-9.82 \pm 0.04) \times 10^{-10}$. 

\subsection{SM prediction of $g-2$ of the muon} \label{g-2muon}

From these results for $a_{\mu}^{\rm had, \, LO \, VP}$ and $a_{\mu}^{\rm had, \, NLO \, VP}$, the SM prediction of the anomalous magnetic moment of the muon is found to be~\cite{Keshavarzi:2018mgv}
\beq \label{amuSMfinal}
a_{\mu}^{\rm SM}  =  (11\ 659 \ 182.04  \pm 3.56) \times 10^{-10} \, .
\eeq
Comparing this with the current experimental measurement results in a deviation of $\Delta a_{\mu} = (27.06 \pm 7.26)\times 10^{-10}$, corresponding to a $3.7\sigma$ discrepancy.

\section{Conclusions}

The uncertainty of $a_{\mu}^{\rm SM}$ is entirely dominated by the hadronic contributions, where below $\sim2$GeV, the estimates of $a_{\mu}^{\rm had,\, VP}$ and the corresponding uncertainties are determined from the experimentally measured cross sections of individual hadronic final states. These measurements are achieved experimentally via direct energy scan or through radiative return. Many experiments have dedicated programmes to precisely measure these final states, meaning that a vast amount of data is now available and that, in some cases, overall precision has reached the sub-percent level. However, data tensions are evident between measurements of the same hadronic channels from different experiments, which reduces the overall quality of the data combinations used to determine $a_{\mu}^{\rm had, \, VP}$. 

The KNT18 analysis has completed a full re-evaluation of the hadronic vacuum polarisation contributions to the anomalous magnetic moment of the muon, $a_{\mu}^{\rm had, \, VP}$. Combining all available $e^+e^- \rightarrow {\rm hadrons}$ cross section data, this analysis found $a_{\mu}^{\rm had, \, LO \, VP} = (693.26 \pm 2.46)\times 10^{-10}$ and $a_{\mu}^{\rm had, \, NLO \, VP} = (-9.82 \pm 0.04)\times 10^{-10}$. This has resulted in a new estimate for the Standard Model prediction of $a_{\mu}^{\rm SM}  =  (11\ 659 \ 182.04  \pm 3.56) \times 10^{-10}$, which deviates from the current experimental measurement by $3.7\sigma$.

\end{document}